\documentclass[aip,10pt,twocolumn]{revtex4}

\usepackage{setspace}
\usepackage{epsfig,latexsym,graphicx,amsfonts,bm,pifont,epstopdf}

\newcommand{\beq}{ \begin{equation} }
\newcommand{\eeq}{ \end{equation} }
\newcommand{\beqs}{ \begin{eqnarray} }
\newcommand{\eeqs}{ \end{eqnarray} }

\begin{document}

\title{Paramecium swimming in capillary tube}
\author{Saikat Jana}\email{jsaikat@vt.edu}
\affiliation{Department of Engineering Science and Mechanics, Virginia Polytechnic Institute and State University,VA 24061, USA}
\author{Soong Ho Um}
\affiliation{Materials Science and Engineering, Gwangju Institute of Science and Technology, Republic of Korea}
\author{ Sunghwan Jung }\email[Corresponding Author    :    ]{sunnyjsh@vt.edu}
\affiliation{Department of Engineering Science and Mechanics, Virginia Polytechnic Institute and State University,VA 24061, USA}

\date{\today}

\begin{abstract}
Swimming organisms in their natural habitat navigate through a wide array of geometries and chemical environments. Interaction with the boundaries is ubiquitous and can significantly modify the swimming characteristics of the organism as observed under ideal conditions.  We study the dynamics of ciliary locomotion in {\it Paramecium multimicronucleatum} and observe the effect of the solid boundaries on the velocities in the near field of the organism. Experimental observations show that {\it Paramecium} executes helical trajectories that slowly transition to straight line motion as the diameter of the capillary tubes decrease. Theoretically this system is modeled as an undulating cylinder with pressure gradient and compared with experiments; showing that such considerations are necessary for modeling finite sized organisms in the restrictive geometries.
\end{abstract}

\maketitle
\section{INTRODUCTION}
Microorganisms use a variety of propulsion mechanisms to swim around in their habitat for predator evasion \cite{Stocker1} or locomotion towards favorable gradients \cite{Berg1}. The locomotory behavior and performance is usually controlled by chemical or hydrodynamic cues arising due to interaction with the local environment. The behavior of organisms can be studied considering them as a single entity or in a group. Collective motion of organisms have revealed vivid characteristics; for example sperm swimming near surfaces execute circular trajectories and aggregation of organisms near the surfaces is an interesting phenomena \cite{Lauga1}. In addition, swimming velocity and direction, tumbling probability and turn angle in capillaries have been characterized \cite{Biondi1,Liu1}. Motility of cells and the morphological changes due to restrictive geometries are active area of interest \cite{Mannik1,Wang1}. In all of these systems organisms often interact with the nearby surfaces of organisms or the boundaries (geometrical constraints) imposed by the nature on the motion; which causes the organism to exhibit varied swimming characteristics as compared to its motion in ideal infinite fluid medium.

In many eukaryotic microorganisms coordinated motion of cilia helps in propagating metachronal waves; which propel the organism\cite{Sleigh1}. Millions of ciliary hairs in mammals help in mucus transport and also function as sensory organelles that help in maintaining balance\cite{Smith1}. The ciliary beat is an interesting physical phenomena and has been studied extensively from biological point of view\cite{Sleigh2}. For example, experiments have been conducted to gain insights into the electro-physiological changes in the ciliary beat \cite {Machemer3} and taxis of ciliates under various conditions have been categorized \cite {Dryl1}. The effect of high viscosities \cite{Machemer1} on the locomotory traits of {\it Paramecium} especially with regards to changes in wave velocities, amplitude of beat of the cilia, wavelength of the ciliary beat have been extensively documented. Experiments on swimming of ciliates in vertically aligned tapered glass tubes \cite {Winet1} have revealed interesting locomotory traits.

A 2D wavy sheet that can be used to model simplified swimming motion in many micro-organisms provided the theoretical framework for modeling microorganisms, by considering small amplitude expansions of the propagating wave\cite{Taylor1}. Further studies into the swimming of 2D sheet surrounded by the planar boundaries revealed propulsive advantages for specific beat patterns \cite{Katz1}. Phase locking in wavy sheets have been studied theoretically \cite{Lauga2} and flow patterns in the near-field of cilia have been investigated numerically \cite{Dauptain1}. Thousands of cilia in {\it Paramecium} beat just out of phase to propagate waves in fluid and hence can either be modeled as infinitely long cylinder \cite{Blake3} or sphere \cite{Blake4} with surface undulations or as a spheroid with slip velocity \cite{Keller1}. These models were also used to validate swimming patterns in variety of ciliates \cite{Brennen2} and to develop a boundary layer theory for predicting the near and far field velocities of ciliary micro-organisms in the unbounded fluid \cite{Brennen1}.

Previous experiments have involved measuring the average velocity; however various trajectories executed by {\it Paramecium} have not been considered. Theoretical studies involving swimming {\it Paramecium} have mostly focussed on infinite models without consideration of the boundary effects. We present a combined experimental and theoretical approach to reveal the locomotive patterns of {\it Paramecium multimicronucleatum} and rationalize the pressure gradient effects, on the swimming behavior in confined spaces. In section \ref{sec:exp}, we explain the experimental methods of introducing the organisms in confined geometries like the capillary tubes and the techniques used to visualize their motion. A theoretical model with a pressure gradient is developed to understand the effect of boundaries in section \ref{sec:the}. In section \ref{sec:res}, we compare the predicted swimming velocity with the experiments and discuss other important parameters that might affect locomotion of microorganisms close to the boundaries in section \ref{sec:disc}. 

\section{EXPERIMENTS} \label{sec:exp}
\subsection{EXPERIMENTAL METHODS}
{\it Paramecium multimicronucleatum} is a single-celled eukaryote commonly found in warmer regions of the freshwater ponds \cite{Sleigh1}. Cultures were reared in a double wheat medium and subcultures were placed every 11 days when they reach their peak population. {\it Paramecium} at the beginning of the their exponential growth curve were used for the experiments.  The cultures were centrifuged and washed twice in a Buffer solution consisting 9 mM CaCl$_2$, 3 mM KCl, 5 mM Tris-HCl (pH 7.2)  to remove the debris and were allowed to equilibrate for 30 min. The equilibrated cells were then observed under the Leica DMI 3000 microscope at 5X, 20X and 40X  with bright field or DIC optics and their motion was recorded using Redlake MotionXtra N3 camera.

The ciliary coordination in ciliates is often controlled by a complex collection of external cues that causes the organism to change the frequency or other parameters of wave propagation. However the difference in directions of the  propagating metachronal wave and swimming direction causes the organism to move in helical path. Broadly the locomotory gait can be classified as forward(anterior) or backward(posterior) swimming; with the forward swimming exhibiting different helical modes. The anterior swimming direction and wave propagation direction are separated by 135$^\circ$ in right handed helix swimmers and by 225$^\circ$ in left handed helical swimmers \cite{Dryl1}. The ciliary reversal modes of locomotion are characterized by little or no helical motion and a coasting motion with lower swimming speed. The helical modes of swimming in contrast to the ciliary reversal modes allows us to better characterize  the change in the locomotive pattern and hence will be used to study the effect of boundaries.

To investigate the flow-field around the organism suspensions of Polystyrene microspheres (5 $\mu$m diameter, Thermo Scientific) prepared in EDC solution were introduced into the cultures. Small volume of the cultures 5 $\mu$l were then placed on the glass slide which created a very thin film and allowed us to visualize the 2D flow field around the {\it Paramecium}. Figure 1(a) shows a {\it Paramecium} swimming in a thin film of liquid. We can see there are two strong vortices forming on the lateral sides of the organism; showing the strong tangential velocity of cilia on the far field of the organism.

\begin{figure}[!htb] 
    \centering
        \includegraphics[width=.5\textwidth]{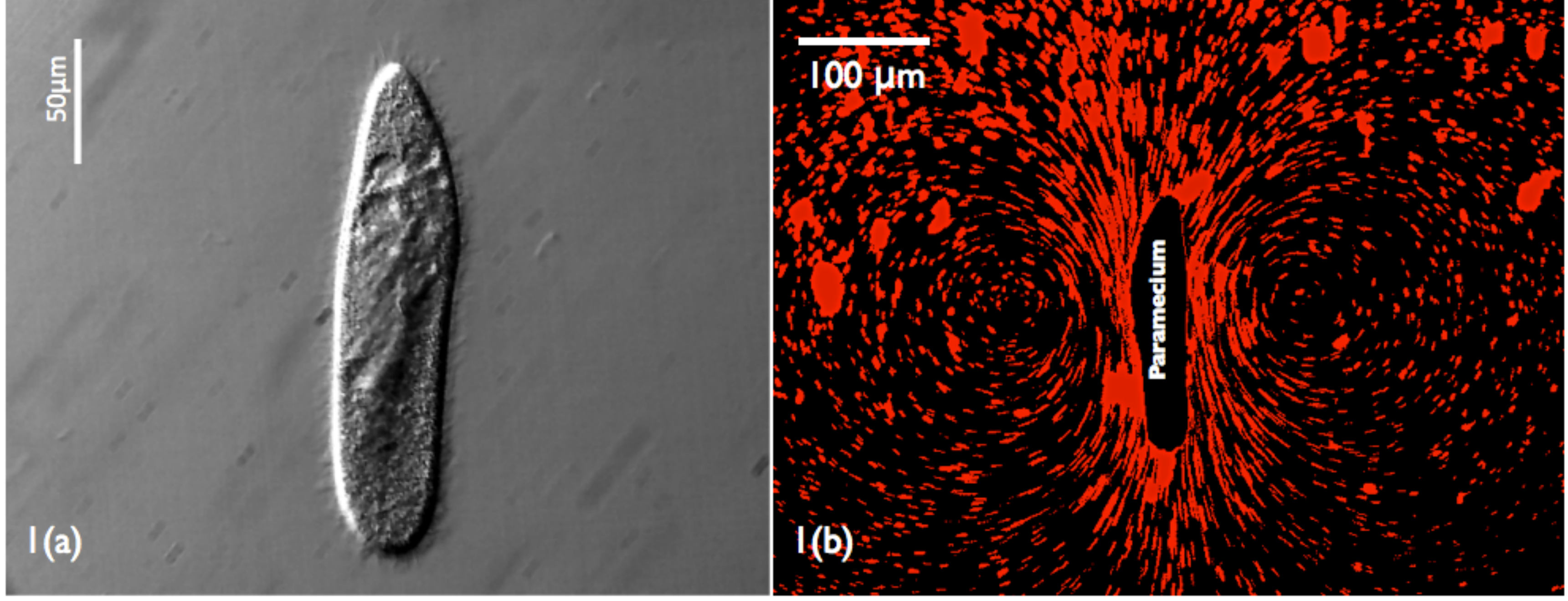}
\caption{(a) A {\it Paramecium} swimming in a thin layer of fluid. (b) Streaklines of particles in flows generated by the cilia}
    \label{fig1}
\end{figure}

The effect of confined geometries on the ciliary dynamics is examined by introducing the organisms in capillary tubes. Tubes required for this purpose are manufactured by attaching a dead weight to the end of the borosilicate glass pipettes and by heating their tips. By controlling the value of the dead weight and the intensity of the applied heat different diameters of capillary tubes ranging from 90$\sim$250 $\mu$m were manufactured. Some commercially available tubes with specific diameters(d=100,150, 200 $\mu$m) made of borosilicate glass were ordered from Vitrotubes. The equilibrated cultures were then transferred to the extruded glass pipettes where they got pulled into the small constant cross section of the tube due to capillary forces.

\subsection{EXPERIMENTAL OBSERVATIONS}

{\it Paramecium} were found to have a long and slender structure with typical lengths around 212$\pm$14 $\mu$m and the width about 57$\pm$5 $ \mu$m (shown in Fig. \ref{fig1}(a)). The velocity of these micro-swimmers in unbounded fluid was found to be 1064$\pm$83 $ \mu$m/s. We then performed experiments with capillary tubes of different diameters. {\it Paramecium} swimming in buffer (isotonic solution) were put into the capillary tubes which caused them to be confined in small circular geometry. A generic code written in MATLAB was used to track the motion of these organisms and the velocities were computed. 

In order to measure the vital parameters for swimming; we captured the cilia motion with the high speed camera at 300 fps, which allowed us to visualize the metachronal wave propagation over the organism. Each cilium was found to be 10$\sim$12 $\mu$m in length and $0.2$ $\mu$m in diameter and beats slightly out of phase compared to the nearby cilium, thereby causing a traveling wave to pass over the surface of the organism. The typical wavelengths of the metachronal waves measured from our experiments were 27 $\mu$m, half peak to peak amplitude 4.2 $\mu$m and the frequency of the beat being around 30 Hz \cite{Sleigh1}. We assume that the cilia are so closely packed that the fluid does not penetrate the material wave, thereby allowing no slip boundary condition to hold good. 

Imaging of round capillaries under microscope caused optical distortions which leads to recording of altered amplitudes and velocities. We directly took images of the cross-section of various capillaries to get a relation between the true and the observed inner diameters of the tubes, which was further used to correct the observed amplitude and velocities of the organisms. 
\begin{figure}[!htb] 
    \centering
        \includegraphics[width=.5\textwidth]{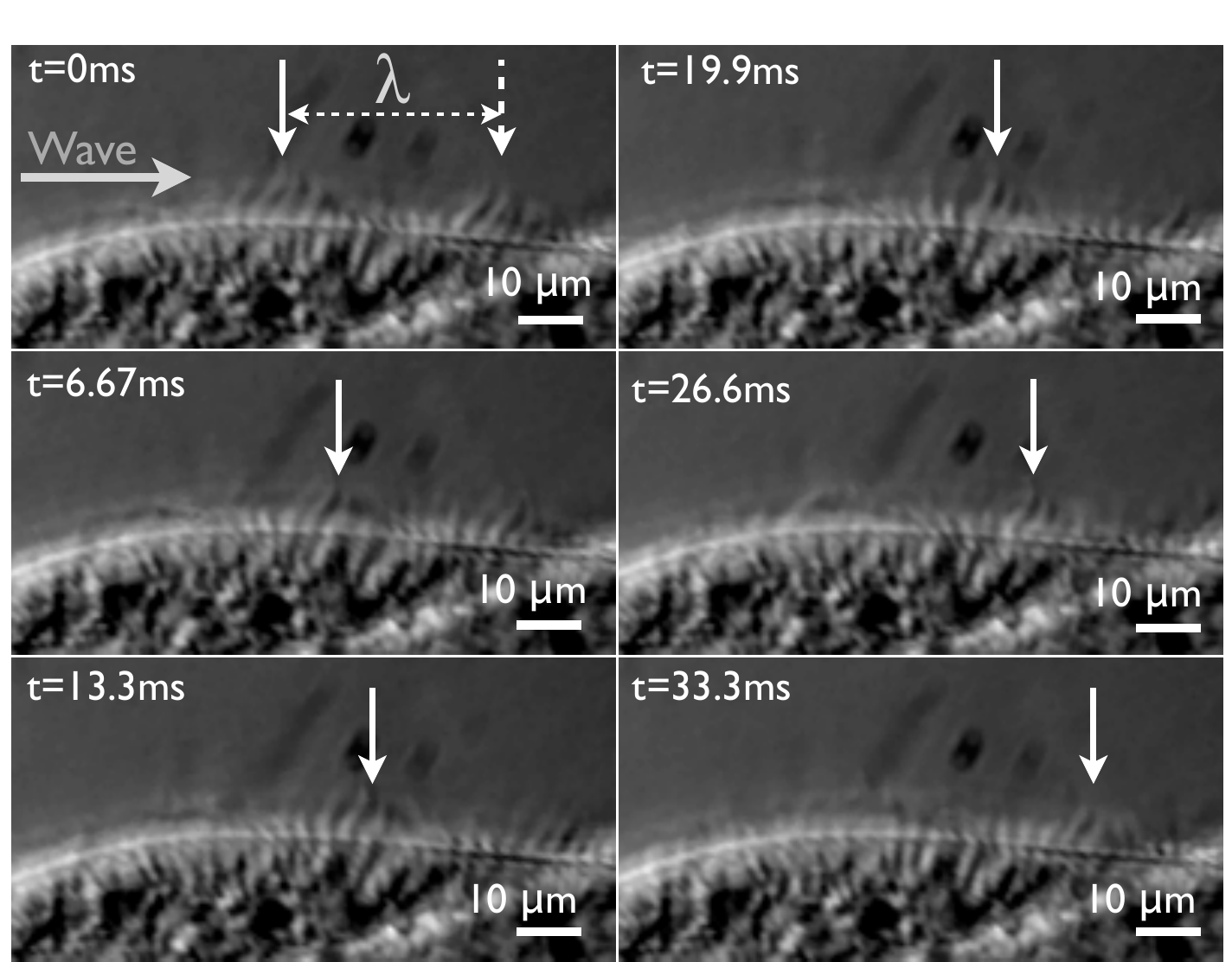}
\caption{The metachronal waves propagated by the cilia of the {\it Paramecium} over a period of 33.3 ms. The sweep of the cilia and the direction of propagation of the wave can be visualized. Arrows follow the peak amplitude of the waves.}
    \label{fig2}
\end{figure}

\begin{figure*} [tb]
    \centering
        \includegraphics[width=.8\textwidth]{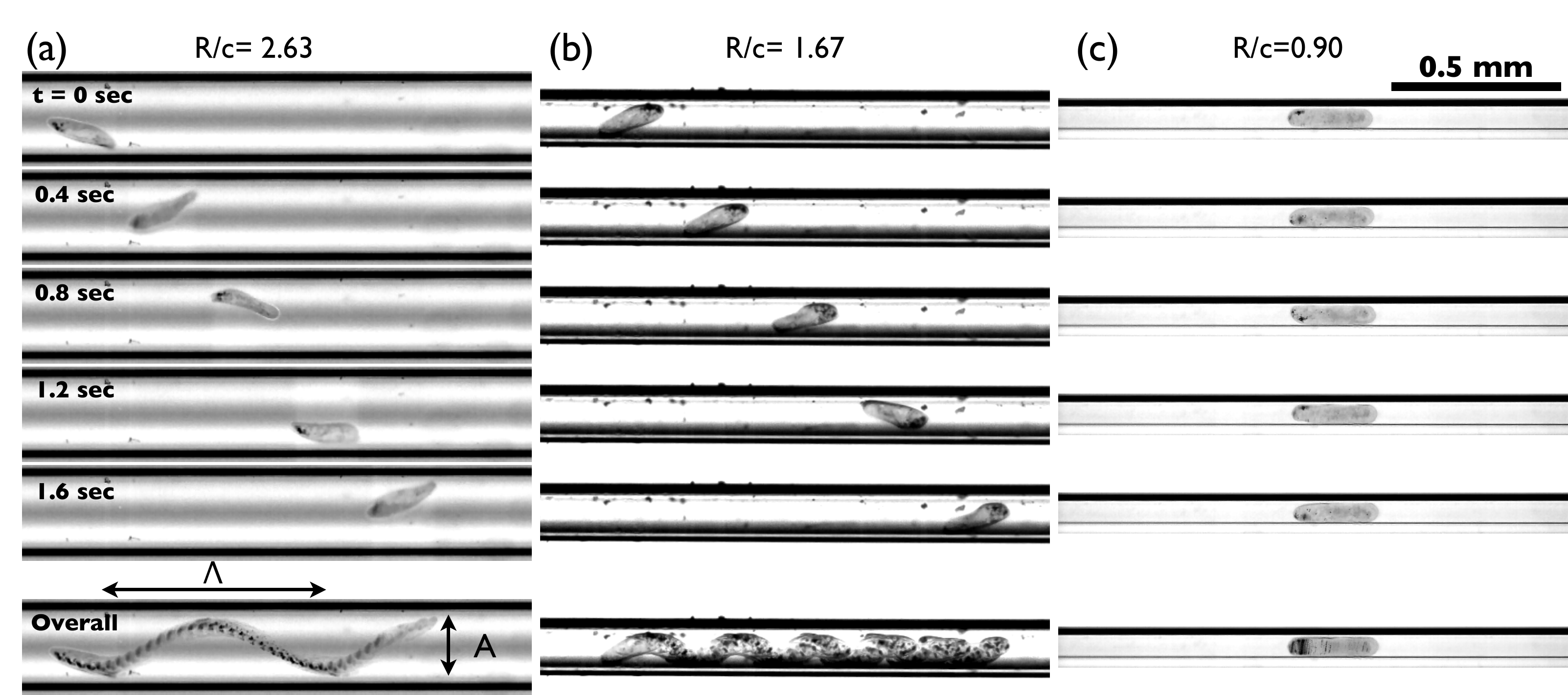}
\caption{Swimming of {\it Paramecium} in tubes of different diameters. Fig. (a) shows swimming in large tube (R/c =2.63) where the trajectory of the motion is helical. In swimming inside tubes of intermediate diameters (R/c=1.67 ) small wavelength helices are seen. In very small tubes(R/c = 0.9) as shown in Fig. (c) {\it Paramecium} swims in a straight line.}
    \label{fig3}
\end{figure*}
In tubes of extremely small diameter the swimming velocity of the organism was very low with almost a straight line motion with variable rotation rates. Whereas in tubes of larger diameter the organism was seen to move in a helical path instead of straight line motion. We applied a correction factor for the path of the swimming organisms, as the image analysis only revealed the 2D projection of the helix.
 
It is observed that the {\it Paramecium} swims slowly as the tube diameter is decreased. This can be attributed to the increased drag felt by the organism due to the proximity of the boundaries. For the capillary tubes whose diameter were very close to the diameter of the {\it Paramecium}, the swimming velocity was close to zero. 

In tubes of smaller diameter ($R/c \sim 1.5$) we observed that a backward (posterior) swimming {\it Paramecium} executed a helical swimming trajectory with small amplitude wavelengths. Such swimming gait have not been reported before \cite{Fukui1,Dryl1}. In this range of tube diameters and for the forward swimming {\it Paramecium}, little or no helical motion of the organism is observed.

\section{THEORETICAL MODEL} \label{sec:the}
\begin{figure}[!] 
    \centering
        \includegraphics[width=.45\textwidth]{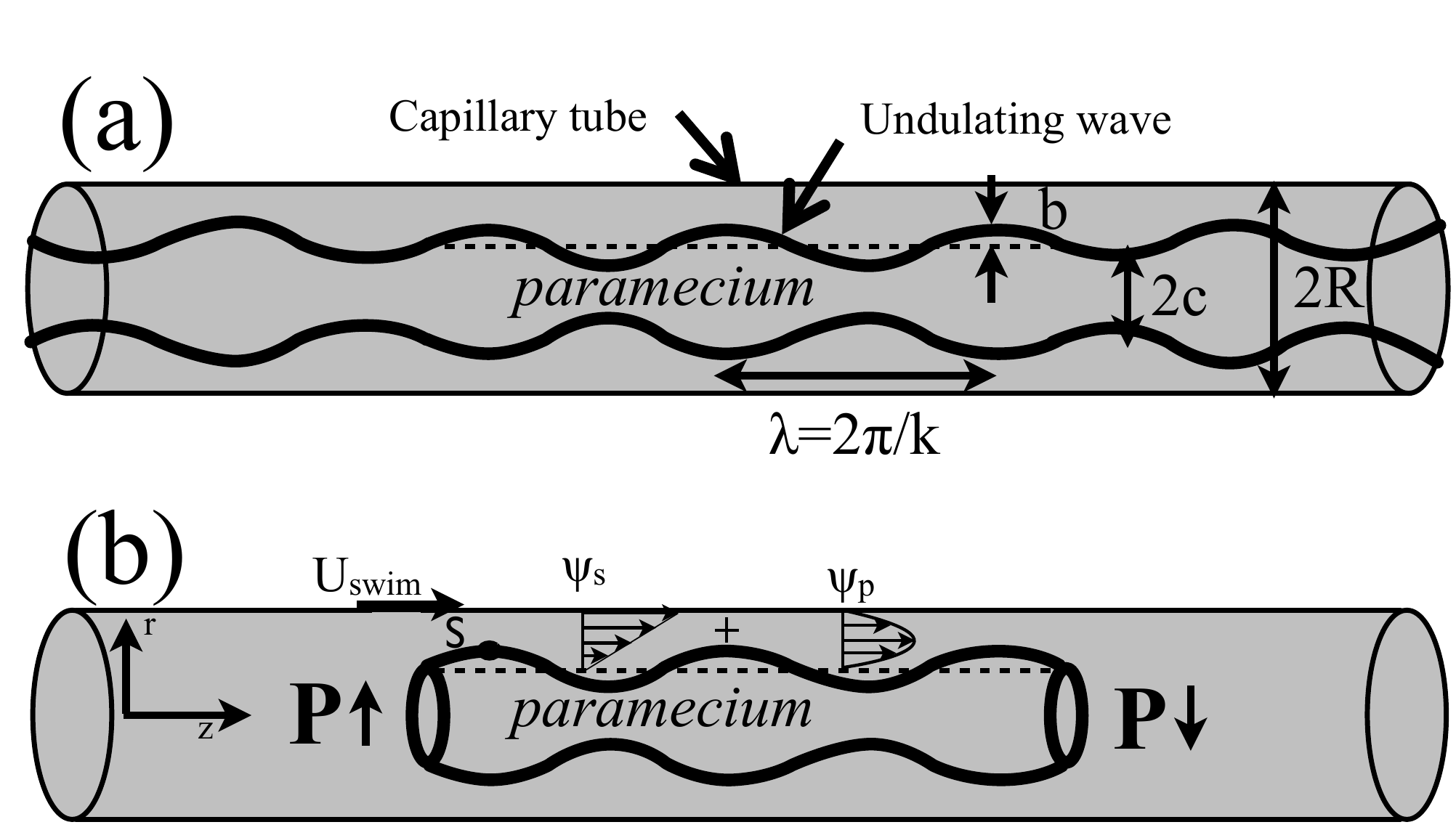}
\caption{Schematic of a wavy Fig.(a) infinite cylinder or Fig. (b) finite cylinder swimming inside the cylindrical tube(boundaries). It is assumed that the organism is swimming in the negative z direction with a velocity $U_{swim}$. By transforming the coordinate system to one, where the organism is swimming with velocity $U_{swim}$ we get the current schematic with the descriptions of the velocities on the walls of capillary. }
    \label{fig4}
\end{figure}

The governing equations for very low Reynolds number flows $(Re\ll1)$ are the Stokes equations: {$\nabla p=\mu \nabla^2{ \bf u}, \nabla\cdot{\bf u}=0$}. Since these organisms have a large length to diameter ratio; they can be effectively modeled as cylinders. As seen in the Figure 2 the cilia create synchronized motion to produce metachronal waves; to an observer this seems like a material wave propagating on the surface of the organism. Under the assumption that no fluid penetrates the wave of cilia tips since they are densely packed, the concept of envelop model can be applied for the physical system. The boundary in this case is the circular capillary tube; thus the problem reduces to modeling a cylinder with a wavy surface swimming inside a cylindrical geometry.

Figure \ref{fig4} shows the schematic of the organism swimming the tube with a velocity $U_{swim}$. {\it Paramecium} has previously been modeled as an infinitely long cylinder so that there is no pressure gradient at the front and the back while swimming in the unbounded fluid. Similar infinite models can be developed for a {\it Paramecium} swimming inside a tube \cite{Blake3}. However, due to the presence of the confined spaces the finite length {\it Paramecium} experiences a pressure gradient at its ends that influences its locomotion within tubes. Considering the propagating wave to be of frequency $\omega$,wavelength $\lambda$, radial amplitude $b$, transverse amplitude $a$, and wave number $k=\frac{2\pi}{\lambda}$; any material point {\bf S} on the undulating surface can be written as: 
\beqs
r_s= c+b \sin(kz-\omega t), z_s=z+a\cos(kz-\omega t+\phi)
\label{eq1}
\eeqs
where $\phi$ is the phase difference.

We work in the frame in which the organism is swimming with a velocity $U_{swim}$. 
In the small amplitude limit, the boundary conditions of the surface of the organism at $r = c$ and on the solid walls at $r=R$ can be written as: 
\beqs
&&u_r|_{r=c}= -b\omega \cos(kz-\omega t), u_z|_{r=c}= a\omega \sin(kz-\omega t+\phi)
\nonumber \\
&& u_r|_{r=R}= 0, u_z|_{r=R}= U_{swim}
\label{eq2}
\eeqs

Thus the swimming problem can thus be envisioned as sum of pressure driven flow and shear flow in the narrow annulus that surrounds the {\it Paramecium}. We seek a solution in terms of the streamfunction $\psi$ such that $\psi= \psi^{(p)}+\psi^{(s)} $ where $\psi^{(p)} $is the streamfunction corresponding to pressure driven flow and $\psi^{(s)} $due to the the shear flow. Using cylindrical co-ordinates and axisymmetric potential theory we can write the velocity components to be $u_r =-\frac{\partial \psi}{r\partial z}$,$u_z= \frac{\partial \psi}{r\partial r}$. Solving for the pressure driven flow with stationary boundaries we can get the streamfunction to be:
\beq
\psi ^{(p)}= \frac{1}{4\mu}\frac{\partial p}{\partial z} \left [ \frac{R^2-c^2}{\ln\frac R c}\left(\frac {r^2} 4-\frac{r^2}{2}\ln \frac r  c\right)+\left(\frac{r^4} 4-\frac {c^2r^2} 2\right)\right]
\eeq

For the shear flow part we can substitute the velocity components in the Stokes equation and by taking curl we end up with a equation of the form  $\left(\frac{\partial^2}{\partial r^2}-\frac{1}{r}\frac{\partial}{\partial r}+\frac{\partial^2}{\partial z^2}\right)^2 \psi^{(s)}=0$.

Rewriting coordinate systems as $ \eta =kr$ and $\zeta= kz-\omega t$ and using separation of variables in $\eta$ and $\zeta$ we can obtain a streamfunction solution of the form:
\beqs
\psi^{(s)}=\frac{U_{swim}\eta^2}{2k^2}
+\sum_{i=0}^{m}{}  F_{n}\sin(\zeta)+\sum_{i=0}^{m}{} G_{n}\cos(\zeta)
\eeqs
where,
\\
$ F_{n}= \eta \left[A_n K_1(n\eta)+B_n\eta K_2(n\eta)+C_n I_1(n\eta)+D_n\eta I_2(n\eta)\right]$
 \\$G_{n}= \eta \left[ \alpha_n K_1(n\eta)+\beta_n\eta K_2(n\eta)+\gamma_n I_1(n\eta)+\delta_n\eta I_2(n\eta)\right]$\\and  $A_n,B_n,C_n,D_n, \alpha_n,\beta_n,\gamma_n,\delta_n$ are the constants to be determined from the boundary conditions and $I$ and $K$ are modified Bessel functions of the first and second kind. 
  
We seek perturbation expansions of the velocities $u_r=-k^2\frac{\partial \psi}{\eta \partial\zeta}$ and $u_z=k^2\frac{\partial \psi}{\eta \partial\eta}$ derived from the stream function to the $O^0(bk)$ and $O^1(bk)$ and compare the velocities with Equation \ref{eq2}. For calculations of the zeroth and first order the $\alpha_n,\beta_n,\gamma_n,\delta_n$ does not have any contribution and hence we are left to determine four unknowns for the problem.

Coupled with the pressure driven flow this gives us a set of five equations and six unknowns $A_1,B_1,C_1,D_1, U_{swim}, \frac{\partial p}{\partial z}$. Also from the flux condition of the swimming {\it Paramecium} we have $Q = U_{swim} \pi c^2 = \int\int u_z r dr d\theta$. This gives us a condition between $U_{swim}$ and $\frac{\partial p}{\partial z}$. Thus we have a set of five equations and five unknowns which can be solved to find out the constants and the swimming velocity $U_{swim}$ of the organism. The expression for swimming velocity of the {\it Paramecium } comes out to be:
\beqs
&&U_{swim}=\frac{k^3b}{2}F_n(\eta_c)+\frac{a^2k\omega}{2}+\frac{abk\eta_c \cos\phi}{8\mu}\frac{\partial p}{\partial \zeta}\nonumber \\
&&-\frac{abk\cos\phi}{8\mu\eta_c}\frac{\partial p}{\partial \zeta} \frac{\eta_{R}^2-\eta^2_{c}}{\ln \frac {\eta_R} {\eta_c}}
\eeqs

The above expression shows that the swimming velocity is directly dependent on $(ak)^2$ and also the pressure gradient terms. For the infinite boundaries case with no pressure gradient the  above equation reduces to the following expression:
\beq
U_{blake} = \frac{\omega k}{2} \left[\frac{(K_0^2-K_1^2)b^2}{K_1^2-K_0K_2}-\frac{2K_1^2ab\cos\phi}{\eta_c(K_1^2-K_0K_2)}-a^2\right]
\eeq
which is same as in\cite{Blake3}. 

\section{RESULTS} \label{sec:res}
\begin{figure}[htd] 
    \centering 
        \includegraphics[width=.45\textwidth]{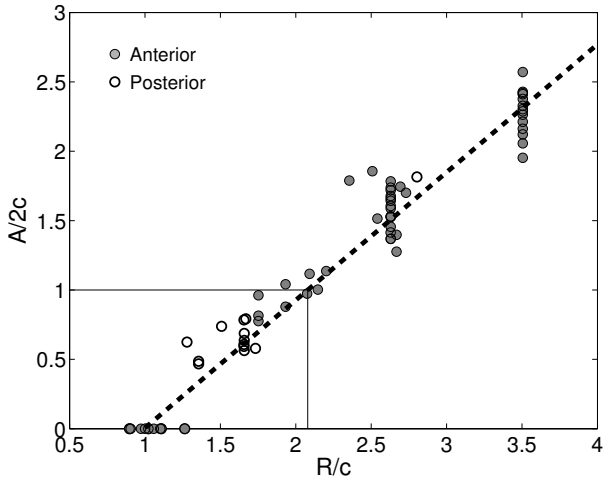} 
\caption{Plot of the non-dimensional amplitude of the helical path vs the non-dimensional radius of the tube. The plot shows a linear variation which increases with the radius of the tube.}
\label{fig5}
\end{figure}

The motivation of the study was to rationalize the behavior of the organisms in close proximity to the boundaries. As the {\it Paramecium} swims inside the tube it traces out a helical path with the length of the body being aligned in the swimming direction. Figure \ref{fig5} shows the variation of the amplitude of the organism as it swims in capillary tube of different diameters. 
Theoretically we considered the small amplitude expansions for the waves and hence the expression derived for the swimming velocity would only be valid for smaller capillary tubes in which there is very less off axis movement. In larger tubes the organism does remain very close to the glass surfaces; but there is asymmetry in the boundary conditions that need to be applied. From the plot above and the constraints on the amplitude of swimming of {\it Paramecium} we can conclude that for $A/(2c)<1$  our experimental and theoretical results would remain valid. This gives us corresponding value of the non-dimensional radius of the capillary tube $R/c < 2$.

\begin{figure}[htd] 
    \centering
        \includegraphics[width=.45\textwidth]{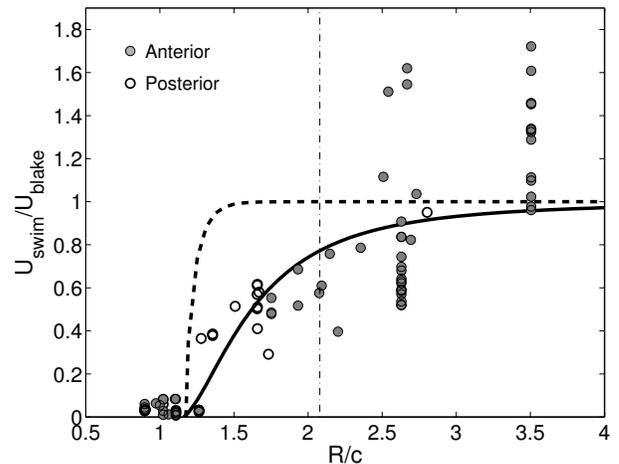}
\caption{Plot of normalized swimming velocity vs. the no-dimensional radius of the tube. Two different cases are compared, one with the dotted line shows the model of an infinite ciliate swimming in capillary tube. The solid line shows the velocity variation of the finite sized {\it Paramecium}  inside confined geometries. Dots represent the experimental result and compares well with the finite pressure gradient model due to the restrictive environment.}
\label{fig6}
\end{figure}

Figure \ref{fig6} shows the variation of velocity with the radius of the capillary tube. The dotted line shows the swimming velocity when no pressure gradient effects are considered and the solution converges very quickly to the case for which the boundaries are at infinity. In the experiments we observe a slowly increasing trend of velocity. For a finite size {\it Paramecium} swimming inside a restrictive geometry there is a finite pressure gradient at the front and the back. The plot of the swimming velocity considering finite pressure gradient is shown by the solid line. It shows a slowly increasing trend and finally converges to the infinite boundary and no pressure gradient case for very large $R/c$ values. 

It can be seen that for $R/c<2$ both the experiments and theory show an increasing trend and predictions match quite well. For the larger $R/c$ we see that the velocities are much larger as opposed to that predicted value for the infinite boundary case. From the experiments we see that the organism follows the surface of the glass capillary as it swims in helical path in tubes of larger diameter. This situation can be thought of as {\it Paramecium} swimming close to a single solid wall and hence explains the observation of larger swimming velocities for tubes of larger diameter; where it constantly swims close to the wall. This reflects the large errors  corresponding to the swimming case in larger tubes.

\begin{figure}[htd] 
    \centering
        \includegraphics[width=.45\textwidth]{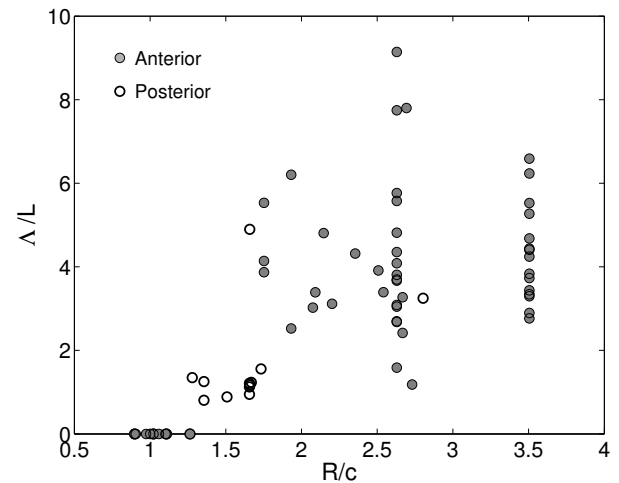}
\caption{Plot of the non-dimensional wavelength vs. non-dimensional radius of the tube. No apparent trend is seen with increasing radius of the tube.}
\label{fig7}
\end{figure}

Figure \ref{fig7} shows the variation of wavelengths of the helical path while they are swimming in the tube. The wavelengths also show a linearly increasing trend for the limit $R/c<2$. In this limit the helices of the swimming are well defined and periodic wavelengths are observed. While swimming in tubes of larger diameter($R/c>2$) it was observed many times that the organism did not execute a full helix.

\section{DISCUSSION}\label{sec:disc}
We investigated the locomotion of ciliary organisms in confined geometries. As the boundaries close on the organism, more viscous effects are felt by the cilia. The swimming velocity of {\it Paramecium} decreases due to the effect of close boundaries. Also in such confined spaces a finite sized organism feels a pressure gradient across the ends. This pressure gradient affects the swimming velocity and needs to be considered while modeling similar self propelling objects in restrictive geometries. Our experiments confirm this observation which show a slowly increasing trend of velocities.

Many interesting questions arise from this study especially about the amplitude of the waves propagated by the {\it Paramecium}. In a restrictive channel the beat of the waves is limited by the dimensions of channel and the size of organism. Due to the proximity, do the amplitudes of the beat change to provide a locomotory advantage?

The helical path traced out by the {\it Paramecium} with the anterior portion of the body aligned towards the local swimming direction is also an interesting locmotory trait. It was also observed that as the radii of the capillary tube increased the radius of the swimming helix also increased. It might be possible that being close to the curved surface provides a propulsive advantage. It was also found that anterior swimming {\it Paramecium} can execute well defined helical paths when put inside capillary tubes of certain diameter. 

The study revealed the interesting locomotory traits in presence of solid wall. Future work would involve investigating the hydrodynamic effects of different textured boundaries on the swimming characteristics of the organism.

\begin{acknowledgements}
The author Soong Ho Um was supported by NCRC program of the Korean Science and Engineering Foundation (Grant No. R15-2008-006-02002-0) and the National Research Foundation of Korea (NRF/MEST) (No. 20100007782; Mid-career Researcher Program). 
\end {acknowledgements}

\end{document}